\documentstyle[draft,aps,epsfig]{revtex}
\def\rxn{$e^+e^-\rightarrow hadrons$}
\def\ftrho{F^{(3)}_\rho}
\def\ferho{F^{(8)}_\rho}
\def\ftomega{F^{(3)}_\omega}
\def\feomega{F^{(8)}_\omega}
\def\ftphi{F^{(3)}_\phi}
\def\fephi{F^{(8)}_\phi}
\begin{document}
\draft
\title{Isospin Breaking Effects in the Extraction of Isoscalar and \\
Isovector Spectral Functions From $e^+e^-\rightarrow hadrons$}
\author{Kim Maltman} 
\address{Department of Mathematics and Statistics, York University, \\
          4700 Keele St., North York, Ontario, CANADA M3J 1P3 \\ and}
\address{Special Research Center for the Subatomic Structure of Matter, \\
          University of Adelaide, Australia 5005}
\maketitle
\begin{abstract}
We investigate the problem of the extraction of the isovector and
isoscalar spectral functions from data on $e^+e^-\rightarrow hadrons$,
in the presence of non-zero isospin breaking.  It is shown that
the conventional approach to extracting the isovector
spectral function in the $\rho$ resonance region, in which
only the isoscalar contribution associated with
$\omega\rightarrow \pi\pi$ is subtracted,
fails to fully remove the effects of the isoscalar component of
the electromagnetic current.
The additional subtractions
required to extract the pure
isovector and isoscalar spectral functions
are estimated using 
results from QCD sum rules.
It is shown
that the corrections are small ($\sim 2\%$) in the isovector case
(though relevant to precision tests of CVC), but very large ($\sim 20\%$)
in the case of the $\omega$ contribution to the isoscalar spectral
function.  The reason such a large effect is natural in the
isoscalar channel is explained, and implications for other
applications, such as the extraction
of the sixth order chiral low-energy constant, $Q$, are discussed.
\end{abstract}

\section{Introduction}
One of the basic ingredients of the Standard Model is the Conserved
Vector Current hypothesis (CVC), which postulates that the charged
(isovector) weak vector current and the neutral isovector component
of the electromagnetic (EM) current are members of the same isovector
multiplet.  Since the charged current spectral function is now measured
rather accurately in $\tau$ decay\cite{aleph97,cleo95,argus87}, CVC
can be tested experimentally, provided, that is, one can extract the isovector
spectral function from data on {\rxn}\cite{gilmanetc,ks90,ei95,ms88,pichetc}.  
In the absence of isospin
breaking, this extraction is straightforward since, for example,
for $n$-pion final states, a state with an even (odd) number of pions
has even (odd) G-parity and hence can be produced only through the
isovector (isoscalar) component of the hadronic EM current.

Isospin breaking, however, complicates the extraction of
both the isovector and isoscalar spectral functions.
Before proceeding, it
is useful to clarify our notation.  We define the standard
$SU(3)_F$ octet of vector currents by 
$J^a_\mu = \bar{q}\gamma_\mu {\frac{\lambda^a}{2}}q$,
where $\lambda^a$ are the usual Gell-Mann matrices.
The electromagnetic current is then
$J^{EM}_\mu = J^3_\mu +J^8_\mu /\sqrt{3}$,
while the scalar correlators, $\Pi^{ab}(q^2)$
(where we will restrict our attention to $a,b=3,8$) are defined by
\begin{equation}
i\int\, d^4x\, \exp (iqx)\, \langle 0\vert T\left(
J^a_\mu (x) J^b_\nu (0)\right) \vert 0\rangle \equiv
\left( q_\mu q_\nu -q^2 g_{\mu\nu}\right) \Pi^{ab}(q^2 ).
\label{scalarcorrelator}
\end{equation}
Defining the spectral functions, $\rho^{ab}(q^2)$, 
corresponding to the $\Pi^{ab}(q^2)$
in the standard manner,
$\rho^{ab}(q^2)={\frac{1}{\pi}}{\rm Im}\, \Pi^{ab}(q^2)$,
the EM spectral function is then 
\begin{equation}
\rho^{EM} = \rho^{33}+{\frac{2}{\sqrt{3}}}\rho^{38}+{\frac{1}{3}}\rho^{88}.
\end{equation}
$\rho^{EM}$ contains the isovector spectral function, $\rho^{33}$, which is
the isospin rotated version of the corresponding charged current
isovector spectral function measured in $\tau$ decays.
Owing to the relation,
\begin{equation}
\sigma (e^+e^-\rightarrow hadrons) = {\frac{8\pi^2\alpha^2}{s}}\rho^{EM}(s),
\end{equation}
however, the cross-section for {\rxn} directly measures $\rho^{EM}$, and not
$\rho^{33}$ or $\rho^{88}$.  If isospin were not explicitly broken, this would 
present no problem since
$\rho^{38}$ would necessarily vanish and, as noted above, one
could in addition identify the states contributing to the isovector ($33$) and
isoscalar ($88$) spectral functions by their G-parity.  
Near threshold it is known, from a study of the mixed isospin
vector correlator ($ab=38$) to two loops in chiral perturbation
theory(ChPT)\cite{km96}, that isospin breaking effects in
$\rho^{EM}$ are negligible.  However, in the region of the $\rho$
and $\omega$ resonances,
isospin breaking is significant, as signalled by
the interference dip in the cross-section
for the $\pi\pi$ final state\cite{barkovetc}.
The conventional method\cite{ks90} for making corrections for this observed
isospin breaking, in order to extract the vector component of the
EM spectral function, is to first parametrize the amplitude
in terms of a sum of $\rho$ and $\omega$ Breit-Wigner resonance
forms (in general one includes also contributions associated
with the higher $\rho$ resonances), and having fitted the
parameters to the observed cross-sections, remove the $\omega$
contribution to the amplitude by hand.  The squared modulus of
the resulting modified amplitude is then used in place of the squared modulus
of the original amplitude to identify that
portion of the cross-section to be associated with the purely
isovector ($33$) portion of the EM spectral function.

The conventional proceedure just described for correcting for isospin breaking,
however, does not, in fact, produce the desired $33$ component of
the vector spectral function.  To understand why this is the case,
let us first define, for the neutral vector mesons, $V=\rho$, $\omega$,
$\phi$, the decay constants $F^a_V$ via
\begin{equation}
\langle 0\vert J^a_\mu \vert V(k)\rangle = m_V F^a_V\, \epsilon_\mu (k)
\end{equation}
where $\epsilon_\mu (k)$ is the vector meson polarization vector,
and $a=3,8$.  $\ftrho$, $\feomega$ and $\fephi$ are non-zero in
the isospin limit, while $\ferho$, $\ftomega$ and $\ftphi$ vanish
in the isospin limit and hence are proportional to
the isospin breaking parameter $\delta m=m_d-m_u$.
In the presence of isospin breaking, all of the neutral vector mesons,
in principle, mix with one another, so the physical states are
of mixed isospin.  If we consider the $\pi\pi$ final state mediated
by the $\omega$ exchange then, to leading order in isospin breaking, this
transition is indeed mediated by the isoscalar component of the
EM current (the intermediate $\omega$ contribution generated by
the isovector current is second order in isospin breaking, one
factor from the coupling $\ftomega$, and one from the isospin
violating $\omega\rightarrow\pi^+\pi^-$ decay vertex), and hence
should be removed if one wishes to extract only the isovector
contributions.  The remaining $\rho$ exchange contributions are,
for a similar reason, however, not purely isovector.  The reason
is that if one considers the $\rho$ exchange contribution to
the EM spectral function, the $38$ component is first order
in isospin breaking (being proportional to $\ftrho\ferho$).  Thus,
the $\rho$ contribution to $e^+e^-\rightarrow \pi^+\pi^-$ necessarily
contains a piece first order in isospin breaking, and associated
with the $38$ part of the EM spectral function, which must be
subtracted in order to isolate the purely isovector $33$
component.  A similar argument shows that there will also be a
$38$ contribution to the measured cross-section for
$e^+e^-\rightarrow \omega\rightarrow 3\pi$, which one would have
to correct for in order to isolate the purely isoscalar $88$
component of $\rho^{EM}$.

In what follows we will discuss how to perform the corrections
associated with first order isospin breaking contributions to the
vector meson decay constants.  The paper is organized as follows.
In section II we first discuss some general issues, which
provide useful qualitative guidelines for the subsequent discussion.
In addition we discuss an analogous case involving the neutral
isoscalar and isovector
axial vector currents, which example serves 
to illustrate the basic features we will
meet in the vector current case of interest, but in a context 
where, unlike the vector case, the isospin breaking decay
constants are already known, being fixed by a next-to-leading order analysis
using Chiral Perturbation Theory (ChPT).  In section III, we then show how,
and with what accuracy, existing QCD
sum rule analyses of the mixed isospin ($38$) vector current
correlator can be used to extract the isospin breaking 
decay constants $\ferho$, $\ftomega$ and $\ftphi$.
In section IV, we use the results of this analysis to evaluate
the corrections required to extract the pure isovector contribution
associated with the $\rho$, and pure isoscalar contributions
associated with the $\omega$ and $\phi$ from the EM cross-section data,
and comment on the effect such corrections would have, for example,
on precision tests of CVC and the extraction of chiral low-energy
constants (LEC's) via the inverse chiral sum rules method\cite{gk95}.

\section{Some Generalities, and an Illustrative Example}

In the Introduction we have explained the reason for the existence of
previously neglected isospin breaking corrections in the extraction
of the isoscalar and isovector spectral functions from {\rxn} data.
We have, however, not yet demonstrated that such corrections can
be expected to be numerically significant.  As will be seen below,
making numerical estimates for the size of these corrections is
non-trivial, and we will be forced to rely on a QCD sum rule 
analysis of the mixed-isospin vector current correlator to make
these estimates.  Since it can be difficult to estimate, in a 
quantitative manner, the errors present in such an analysis, it
is useful to consider any qualitative constraints, based on
general principles, which one might use to judge the plausibility
of the resulting solutions.  We will discuss below the existence
of such constraints based on the framework of effective chiral
Lagrangians.  We will also consider, as an illustrative example,
an analogous case in which the vector currents of the problem at
hand are replaced by axial vector currents.  The advantage of this
example is that the isospin breaking pseudoscalar decay constants
(anologous to the isospin breaking vector meson decay constants
to be determined by the QCD sum rule analysis below) can, in this
case, be computed with good accuracy using the methods of ChPT.
This allows us to show explicitly, in a context where the numerical
accuracy of the evaluation of the isospin breaking corrections
is not open to question, that such isospin breaking corrections
can play a significant role in the correct extraction of 
flavor-diagonal spectral contributions from the spectral
functions of correlators involving
mixed-flavor currents.  Certain features of
the relation of the relative signs and magnitudes of the corrections
in the isovector and isoscalar cases, which recur in the vector
channel, will also be exposed, again in a context where the accuracy
of the numerical estimates is not open to question.  From this 
example we will be able to unambiguously conclude that isospin breaking
corrections of the type also present in the vector channel {\it must} be
expected to be numerically significant, especially for observables
related to differences of weighted integrals of the isovector and
isoscalar spectral functions, for which there will be cancellation
between flavor-conserving contributions, but coherence between the
isospin breaking corrections.

Let us turn then to the qualitative guidance offered,
in the vector channel, by the framework
of effective chiral Lagrangians.  Certain
qualitative features of mixing in the vector meson sector 
follow immediately from the properties of an effective chiral Lagrangian such
as would be obtained by the Callan, Coleman, Wess, Zumino\cite{cwz,ccwz}
construction\cite{EGPdR,drv}.  Note that the
lowest order term in quark masses and derivatives which
produces isospin mixing in the vector meson propagator
matrix involves a single power of the
quark mass matrix (where,
as per the usual chiral counting, external momenta are counted
as ${\cal O}(q)$ and quark masses as ${\cal O}(q^2)$).  Moreover,
at leading order, the isospin violating decay constants 
associated with the isospin pure states vanish.
As a result, at this order, the transformation from the original
isospin pure basis to the physical, mixed isospin basis, is
a rotation, and the isospin breaking decay constants of the
physical particles result purely from the mixing in the physical
states.  If we concentrate on $\rho$--$\omega$ mixing, this would
mean that $\ferho$ and $\ftomega$ were equal in magnitude and
opposite in sign.  If we consider effective isospin breaking
operators higher order in the quark mass/derivative expansion,
new effects come into play.  First, one finds non-vanishing
``direct'' isospin violating couplings of the external vector
currents to the isospin pure states from terms involving both
derivatives and one power of the quark mass matrix.  Second,
terms involving both derivatives and one power of the quark
mass matrix can produce off-diagonal mixing elements in the
wavefunction renormalization matrix, a consequence of which
is that the transformation from the isospin pure to physical
basis is no longer a rotation, but rather the product of a
symmetric matrix and a rotation.  Third, terms with two powers
of the mass matrix will produce modifications in the 
momentum-independent mixing terms in the vector meson propagator
matrix.  One would expect the effect of such higher order terms
to be manifest in deviations from leading order relations such
as $\ferho = -\ftomega$.  The size of such deviations should
be typical of next-to-leading order corrections in $SU(3)_L\times SU(3)_R$,
and hence might be as large as $\sim 30\%$.

We next turn to the illustrative example mentioned above, in which we 
replace the vector mesons and vector currents with
pseudoscalar mesons and axial vector currents.  Not only can
several of the qualitative points just made be
clearly illustrated in this case, but the actual numerical values
of the relevant isospin breaking corrections can, for this example,
be calculated to good accuracy using the techniques of ChPT, since
all of the relevant decay constants are
known at next-to-leading order in the chiral expansion.  This is
in contrast to the case of
the vector current spectral functions, where we will be forced
to rely on a QCD
sum rule analysis to obtain the isospin breaking decay constants.
Let us, therefore, define the axial current combination
\begin{equation}
A_\mu \equiv A^3_\mu +{\frac{1}{\sqrt{3}}}A^8_\mu
\end{equation}
where $A^{3,8}_\mu$ are the $3$, $8$ members of the usual axial
current octet.  We then define the pseudoscalar decay constants, $f^a_P$,
for $P=\pi^0$, $\eta$ and $a=3,8$, via
\begin{equation}
\langle 0\vert A^a_\mu \vert P(k)\rangle = i\, f^a_P \, k_\mu .
\end{equation}
At leading (second) order in the chiral counting the physical
$\pi^0$, $\eta$ basis is a pure rotation of the isospin pure
octet basis $\pi^3$, $\pi^8$
\begin{equation}
\pi^0 = \pi^3 +\theta_0\, , \pi^8\qquad \eta =\pi^8-\theta_0\, \pi^3
\end{equation}
where $\theta_0 =\sqrt{3} (m_d-m_u)/4(m_s-\hat{m})$, with
$\hat{m}=(m_d+m_u)/2$, is the leading order mixing angle, and
the isospin breaking decay constants are produced purely
due to the ``wrong-isospin'' admixture present in the physical
$\pi^0$, $\eta$ states,
\begin{equation}
f^{(8)}_\pi =\theta_0 F\, , \qquad f^{(3)}_\eta=-\theta_0 F
\end{equation}
where $F$ is a second order LEC, identical to both $f^{(3)}_\pi =f_\pi$, the
physical pion decay constant, and $f^{(8)}_\eta = f_\eta$, the
physical $\eta$ decay constant, at this order in the chiral expansion.
When one considers the full effective Lagrangian at next-to-leading
order, however, one finds that (1) there is an infinite renormalization
required to regularize the mixing of $\pi^3$, $\pi^8$ even if
one works with the versions of the fields renormalized in the
isospin limit; (2) there is indeed an off-diagonal
element produced in the wavefunction renormalization matrix and
(3) there are indeed ``direct'' isospin breaking meson-current
vertices.  (For a detailed exposition of the above features
beyond leading order see Ref.~\cite{km95}.)  The effect of all
these features is to produce the following results for the
isospin breaking decay constants\cite{gl85}, 
written in terms of the next-to-leading order expressions for
the $\pi^0$ and $\eta$ decay constants, which differ at this
order in the chiral expansion,
\begin{eqnarray}
f^{(8)}_\pi &=& \epsilon_1 f_\pi \nonumber \\
f^{(3)}_\eta &=& -\epsilon_2 f_\eta .
\label{axf}
\end{eqnarray}
In Eq.~(\ref{axf}), the quantities
$\epsilon_{1,2}$ differ by terms which are next-to-leading order
in the chiral expansion
(the complete expressions can be found in Ref.~\cite{gl85}) and,
using the observed experimental ratio of $f_K/f_\pi$, the
next-to-leading order expressions for the isospin conserving decay constants
imply $f_\eta /f_\pi \simeq 1.3$\cite{gl85}.  

We are now in a position to consider the analogue of the vector current
case of interest.  To this end we imagine that we would like to
obtain the $\pi^0$ and $\eta$ contributions to the isovector and
isoscalar axial spectral functions of the scalar
correlators, $\Pi^{33}_1(q^2)$
and $\Pi^{88}_1(q^2)$, defined by
\begin{equation}
\Pi_{\mu\nu}^{ab}= i\int\, d^4x\, e^{iqx}
\langle 0\vert T(A^a_\mu (x)A^b_\nu (0))\vert 0\rangle
\equiv \Pi^{ab}_1(q^2)q_\mu q_\nu +\Pi^{ab}_2(q^2)
\left( q^2g_{\mu\nu}-q_\mu q_\nu\right)\ . \label{axcorr}
\end{equation}
(We concentrate on the scalar correlators,
$\Pi_1^{ab}$, since it is these correlators which contain the
pole contributions analogous to those of the vector mesons in the
vector current correlators.)  It is straightforward, in this case,
to simply evaluate the 
contributions to the spectral functions, $\rho_1^{ab}$.
Keeping only terms up to first order in isospin breaking and
to next-to-leading order in the chiral expansion, one finds that
\begin{eqnarray}
\rho^{33}_1 (q^2)&=& \left[ f^{(3)}_\pi\right]^2\, \delta (q^2-m_\pi^2)
\nonumber \\
\rho^{38}_1 (q^2)&=&f^{(3)}_\pi f^{(8)}_\pi\, \delta (q^2-m_\pi^2)
+f^{(3)}_\eta f^{(8)}_\eta\, \delta (q^2-m_\eta^2) \nonumber \\
\rho^{88}_1 (q^2)&=& \left[ f^{(8)}_\eta\right]^2\, \delta (q^2-m_\eta^2).
\label{axspec}
\end{eqnarray}
Let us imagine, however, that the way we had to go about extracting
these contributions was by analyzing the ``experimental'' spectral function
$\rho_1^A(q^2)$ of the scalar correlator, $\Pi_1^A$, appearing in
the analogue, $\Pi_{\mu\nu}^A$,
of the correlator of the product of
two EM currents, i.e.,
\begin{equation}
\Pi_{\mu\nu}^A= i\int\, d^4x\, e^{iqx}
\langle 0\vert T(A_\mu (x)A_\nu (0))\vert 0\rangle
\equiv \Pi^A_1(q^2)q_\mu q_\nu +\Pi^A_2(q^2)
\left( q^2g_{\mu\nu}-q_\mu q_\nu\right)
\ . \label{axemcorr}
\end{equation}
The analogue of the standard vector current extraction of the isovector
component would then consist of identifying the $\pi^0$ pole contribution
with the isovector $\rho^{33}_1$ component of $\rho^A_1$ and
the $\eta$ pole term with the isoscalar $\rho^{88}_1$
component thereof.  We can see,
however, that this identification is incorrect.  Indeed, the $\pi^0$
contribution to $\rho^A_1$, to first order in
isospin breaking, is straightforwardly found to be, not $\rho^{33}_1$, but
\begin{equation}
\left[ \rho^A_1\right]_\pi = \rho_1^{33}+{\frac{2}{\sqrt{3}}}
\left[ \rho_1^{38}\right]_\pi
=\Biggl( 
\left[ f^{(3)}_\pi\right]^2 +{\frac{2}{\sqrt{3}}} f^{(3)}_\pi f^{(8)}_\pi 
\Biggr)\, \delta (q^2 -m_\pi^2).
\label{pipole}\end{equation}
Similarly, the $\eta$ pole contributions to $\rho^A_1$, to first
order in isospin breaking, is not
$\frac{1}{3} \rho^{88}_1$ but
\begin{equation}
\left[ \rho^A_1\right]_\eta = {\frac{1}{3}}\rho_1^{88}+
{\frac{2}{\sqrt{3}}}\left[ \rho_1^{38}\right]_\eta
=\Biggl( {\frac{1}{3}}\left[ f^{(8)}_\eta\right]^2 
+{\frac{2}{\sqrt{3}}} f^{(3)}_\eta f^{(8)}_\eta \Biggr) \delta (q^2 -m_\eta^2).
\label{etapole}\end{equation}
Thus, to ``extract'' the isovector and isoscalar spectral functions
from the ``experimental'' spectral function one would actually have
to know the isospin violating decay constants and then form the
combinations
\begin{eqnarray}
\rho^{33}_1 &=& \left[ \rho^A_1\right]_\pi 
- {\frac{2}{\sqrt{3}}} f^{(3)}_\pi f^{(8)}_\pi \, \delta (q^2-m_\pi^2)
\nonumber \\
\rho^{88}_1 &=& 3\left[ \rho^A_1\right]_\eta -2\sqrt{3} 
f^{(3)}_\eta f^{(8)}_\eta\, \delta (q^2-m_\eta^2) .
\label{isoaxcorr}
\end{eqnarray}
Note that, because of the structure of the original current, the
size of the isospin breaking correction in the isoscalar case is 
naturally larger by a factor of $3$ than that in the
isovector case.  If we work
with the values of $\epsilon_{1,2}$ from Ref.~\cite{gl85}
(which correspond to the value of the isospin breaking mass
ratio $r=(m_d-m_u)/(m_d+m_u)$ used in the sum rule analysis
we will employ below)
\begin{equation}
\epsilon_1 =1.37\times 10^{-2}\, ,\qquad \epsilon_2=1.11\times 10^{-2}
\end{equation}
we find that that the ``experimental'' pion pole contribution to
$\rho^A_1$ is a factor of $1+2\epsilon_1 /\sqrt{3}=1.016$ larger than
the true isovector spectral function, and that the {\it nominal}
``experimental'' isoscalar spectral function, obtained by taking
the $\eta$ pole contribution and multiplying by $3$, is smaller
than the true isoscalar spectral function by a factor of
$1-2\sqrt{3}\epsilon_2=0.961$.  Thus, to extract the true
isovector and isoscalar spectral functions, one would have to
multiply the nominal ``experimental'' ones by $0.984$ and
$1.040$, respectively.  Note that the corrections go in the
opposite direction for the two cases and that the magnitude of
the correction is significantly larger in the isoscalar case.
The reason for the latter feature of the results has already
been explained.  The reason for the former is that the leading
order result that the $\pi^0$ and $\eta$
contributions to the $38$ spectral function differ in
sign, but not in magnitude, is still approximately satisfied
by the next-to-leading order expressions, as expected based
on the general arguments above.  Note also that the effect
of these corrections can be greatly enhanced if one considers
combinations which would
vanish at leading order in the chiral counting.  As an example,
if we consider the integral over the $88$-$33$ spectral function,
the nominal value (without the above corrections) is 
$0.69 f_\pi^2$, while the actual value (after corrections) is
$0.77 f_\pi^2$.  The isospin breaking corrections thus
represent a $12\%$ increase for this quantity, much larger than
one would guess based on the typical few percent size of
familiar isospin breaking corrections.

We will see in what follows that many of the features of the
axial current example are recapitulated in the vector current case.
In fact, because in the limit in which one
considers the vector meson multiplet to be ideally mixed,
but takes the decay constants to be otherwise determined
by $SU(3)_F$ symmetry, the $\omega$ EM decay constant is a 
factor of $3$ smaller than the $\rho$ EM decay constant
(rather than just the factor of $\sqrt{3}$ difference
between the ``A'' decay constants of the $\pi^0$ and
the $\eta$ in the above example), the discrepancy between
the size of the corrections in the isovector and isoscalar
channels would be expected to be even greater in the vector case.
We will see below that this expectation is indeed borne out.

\section{The QCD Sum Rule Extraction of
the Isospin Breaking Vector Meson Decay Constants}

Since the $3$ and $8$ components of the vector current octet 
occur in the standard model only in the combination $J^{EM}_\mu$,
it is not possible to directly determine the isospin breaking
vector meson deay constants experimentally.  One can, however, obtain indirect
access to these quantities via a QCD sum rule analysis
of the vector current correlator $\Pi^{38}$.  

The basic idea of such a QCD sum rule analysis\cite{svz,rryrev,narisonbook} 
is straightfoward.
From the behavior of $\Pi^{38}(q^2)$ as $q^2\rightarrow\infty$
in QCD, it is known that $\Pi^{38}$ satisfies an unsubtracted
dispersion relation.  This dispersion relation allows one to relate
the dispersion integral over hadronic spectral function $\rho^{38}$
to the value of $\Pi^{38}$ at large spacelike value of $q^2=-Q^2$.
The latter can be expressed in terms of vacuum condensates using
the operator product expansion (OPE), while the hadronic spectral
function depends on the isospin breaking and isospin conserving
decay constants of the various vector mesons.  The utility of
this relation is greatly enhanced, as first noted in
Ref.~\cite{svz}, if the original dispersion relation is Borel
transformed since, in that case, the higher $s$ portions of the transformed
hadronic spectral integral are exponentially suppressed, while
the contributions of higher dimensional operators on the OPE side
are simultaneously factorially suppressed.  In favorable circumstances
one is then able to write the dispersion relation in a form in
which the parameters of a small number of resonances dominate the
hadronic side, and the contributions of a small number of vacuum
condensates of low dimension operators (which condensates can be
determined from other sum rule analyses) dominate the OPE side.
One can then use the known values of the vacuum condensates to
extract the (in our case unknown) resonance parameters.
Note that it is far preferable, in the case of interest to us,
to investigate the isospin breaking decay constants by analyzing
an isospin breaking correlator, rather than by looking at the
isospin breaking corrections to a correlator which also has
an isospin conserving piece.  This is because of the fact that,
in the latter case, one would have to consider isospin breaking
not only in the vector meson decay constants, but also in
the continuum thresholds, which would be required in the spectral
ans\"atze in order to model the higher $s$ portions of the
hadronic spectral function.  Since that portion of the spectral
model is a rather crude representation of the actual continuum,
this would introduce potentially large, and difficult to control, 
uncertainties into such an analysis.

Let us turn then to the sum rule analysis of the correlator
$\Pi^{38}(q^2)$.  We will, in fact, use the results of existing
analyses\cite{svzro,hhmk,krmsr,ijl}
of the related correlator, $\Pi^{\rho\omega}(q^2)$, defined by
\begin{eqnarray}
\Pi^{\rho\omega}_{\mu \nu}(q)&=&i\int\, d^4x\, \exp (iq\cdot x) \langle
0|T J^\rho_\mu (x) J^\omega_\nu(0)|0\rangle\ \nonumber \\
&\equiv& \left( q_\mu q_\nu -q^2 g_{\mu\nu}\right)\, \Pi^{\rho\omega}(q^2),
\label{relatedcorrelator}
\end{eqnarray}
where $J^\rho_\mu=(\overline{u}\gamma_\mu u-\overline{d}\gamma_\mu d)/2$ and
$J^\omega_\nu=(\overline{u}\gamma_\mu u+\overline{d}\gamma_\mu d)/6$.
This is possible because, truncating the OPE at
${\cal O}(m_q),\ {\cal O}(\alpha_s),\ 
{\cal O}(\alpha_{EM})$ and operators 
of dimension $6$ 
(with either the vacuum saturation hypothesis, or the
rescaled version thereof, for the four-quark condensate contributions),
the $\bar{s}s$ portion of $J^8_\mu$ does not contribute to the OPE 
representation of $\Pi^{38}(q^2)$.  One then has
\begin{equation}
\left[ \Pi^{38}(q^2)\right]_{OPE}=\sqrt{3}\, \left[ 
\Pi^{\rho\omega}(q^2)\right]_{OPE},
\end{equation}
where, 
after Borel transformation (indicated by the operator ${\cal B}$) of
the truncated OPE expression, one finds\cite{svzro}
\begin{equation}
{\cal B}\left[\Pi^{\rho\omega}(s) \right]_{OPE}
={\frac{1}{12}}\left[ c_{0}\,M^2+c_1+{c_2\over M^2}+{c_3\over M^4}\ \right],
\label{borelope}
\end{equation}
with $M$ the Borel mass,
\begin{eqnarray}
c_0&=&\frac{\alpha_{EM}}{16\pi^3} \nonumber \\
c_1&=&{\cal O}(m_q^2)\sim 0\nonumber \\
c_2&=&4\left( {m_u-m_d (1+\gamma)\over 2+\gamma}\right)\langle\overline{q}q
\rangle_0 \nonumber \\
c_3&=&{224\pi\,\gamma\over 81}\alpha_s \kappa\,\langle\overline{q}q
\rangle_0^2
-{28\pi\over 81}\alpha_e \kappa\,\langle\overline{q}q
\rangle_0^2\ ,
\label{wilsoncoefs}
\end{eqnarray}
where $\gamma\equiv \langle\overline{d}d
\rangle_0/\langle\overline{u}u \rangle_0-1$, $\langle\overline{q}q
\rangle_0\equiv (\langle\overline{u}u \rangle_0+\langle\overline{d}d
\rangle_0)/2$, $\alpha_e$ and $\alpha_s$ are the electromagnetic
and strong coupling constants, respectively, and $\kappa$ is the
parameter describing the deviation of the four-quark condensates
from their vacuum saturation values.

The hadronic spectral function, $\rho^{38}$, is parametrized as
\begin{equation}
\rho^{38}(s)={\frac{1}{4\sqrt{3}}}\left[ f_\rho \delta_\rho (s)
-f_\omega \delta_\omega (s)+f_\phi \delta_\phi (s) +\cdots \right]
\label{hadronicspec}
\end{equation} 
where
\begin{equation}
\delta_V(s)={\frac{1}{\pi}}{\frac{m_V\Gamma_V}{(s-m_V^2)^2+m_V^2
\Gamma_V^2}}
\label{bw}
\end{equation}
(which reduces to $\delta (s-m_V^2)$ in the narrow width approximation),
and we have followed the notation and sign conventions employed in
the literature for the parameters $f_V$ describing the strengths
of the various vector meson contributions to the spectral function
in question.  Note that, in the notation introduced above,
\begin{equation}
\frac{f_V}{4\sqrt{3}}=\pm F^{(3)}_V F^{(8)}_V
\end{equation}
where the upper sign holds for $V=\rho$, $\phi$, $\cdots$,
and the lower sign for $V=\omega$.  

A few comments are in order concerning the form of Eq.~(\ref{hadronicspec}).
The first concerns the presence of the 
$\phi$ contribution, which was not included
in the earliest sum rule analyses\cite{svzro,hhmk}.  Recall that we expect,
as a leading order result, that $f_\rho \simeq f_\omega$ (the 
absence of a minus sign in this relation is a consequence of
the sign convention for the definition of $f_\omega$).  This means that,
since from far in the spacelike region the
$\rho$ and $\omega$ masses appear essentially the same, one must expect
significant cancellation between the $\rho$ and $\omega$
contributions to the original dispersion relation for large values of
$Q^2$.  This is especially true in the narrow width approximation.
Based on this observation, it was realized, in Ref.~\cite{krmsr},
that, although $f_\phi$ could be expected
to be significantly smaller than $f_\rho$ or $f_\omega$ (of order
$6$--$7\%$ if one makes an estimate based on the 
Particle Data Group evaluation of the
deviation of vector meson mixing from ideal mixing\cite{krmsr}),
the contribution of the $\phi$ to the actual sum rule need not
be negligible.  The sum rule was then re-analyzed, including the
$\phi$ contribution, though still in the narrow width approximation.  
The results supported the qualitative arguments
regarding the importance of including the $\phi$ contribution, and
simultaneously cured a physically unpleasant feature of the earlier analyses,
in which contributions from the $\rho^\prime$-$\omega^\prime$
region of the spectral function were as important, or more important,
than those from the $\rho$-$\omega$ region in determining, through
the original dispersion relation,
the value of the correlator near $q^2=0$.  It was, however, then
subsequently pointed out\cite{ijl} (IJL), again because of the high degree of
cancellation between $\rho$ and $\omega$ contributions 
to the dispersion integral in the far spacelike
region, that the use of the
narrow width approximation for the $\rho$ might also be a rather poor
one.  This was borne out by the numerical analysis
of Ref.~\cite{ijl}.  Of particular note is the fact that, introducing
the $\rho$ width into the spectral ansatz for the $\rho$ contribution,
one finds that $f_\rho$ and $f_\omega$
are lowered in magnitude by a factor of $\sim 6$, and $f_\phi$ by a
factor of  $\sim 2$ compared to the values extracted using the
narrow width analysis. 
With the above understanding in mind, we
will, therefore, in what follows, employ the spectral ansatz 
of Eq.~(\ref{hadronicspec}) in the form
arrived at in Ref.~\cite{ijl}, i.e., with the $\omega$ and
$\phi$ treated in the narrow width approximation, but
the $\rho$ treated using the Breit-Wigner form given in Eq.~(\ref{bw}).

It should be stressed that, in using the results of Ref.~\cite{ijl},
the present analysis
relies {\it only} on the extraction of the hadronic spectral function for the
mixed-isospin vector current correlator, accomplished in that reference
using QCD sum rules, and not on the attempt by the authors of
Ref.~\cite{ijl} to interpret this spectral function in terms of
off-shell vector meson propagators.  The latter interpretation (and
that of Ref.~\cite{hhmk}) is
necessarily incorrect, since off-shell Green functions are well-known
to be altered by redefinitions of the hadronic fields and, as such, are
not capable of being related to physical objects such as the 
mixed-isospin vector current correlator (an earlier claim by the
present author, contained in
Ref.~\cite{krmsr}, that the rescaled versions of the vector fields
represented a possible field choice for the vector mesons is also
incorrect).  The only questions relevant to the use of the
spectral function solution of Ref.~\cite{ijl} in the present work
are then (1) is the sum rule reliable for the scales at which
it is employed and (2) is the original ansatz for the spectral
function plausible in form.  While it is not possible to provide
a rigorous proof of the suitability of the analysis of Ref.~\cite{ijl},
there is considerable indirect evidence in its favor.  First, although
a resonance-saturation ansatz involving s-dependent widths, 
but constant real parts of the resonance self-energies, does 
not rigorously implement
the constraints of analyticity and unitarity, it is well-known from
phenomenological studies of {\rxn} that such ans\"atze, nonetheless,
can provide fits of very high numerical accuracy to the experimentally
measured spectral function (see Ref.~\cite{vmd} for a recent detailed
discussion).  This is also true of the resonance-saturation fits
to the timelike pion form factor measured in hadronic $\tau$
decays mediated by the isovector current (where, for example,
the naive form of the resonance contributions involving s-dependence
only in the widths, produces a fit of even slightly better quality
than does an ansatz employing the Gounaris Sakurai form, which
correctly implements analyticity and unitarity -- see the results
of Table 3 of Ref.~\cite{aleph97}).  Moreover, a recent analysis
of the isovector (isospin conserving) vector correlator using a 
continuous family of
finite energy sum rules
as a means of implementing the dispersion relation
implicit in QCD sum rules shows that
fitting a naive form of the resonance saturation ansatz for
the hadronic spectral function to the OPE representation (where
the widths and decay constants of the resonances are free parameters
determined by the fit to the OPE) reproduces the experimentally
measured spectral function\cite{aleph97} to an accuracy of a few percent
and, moreover, requires a phenomenological value of the $\rho$ width within a
few MeV of that obtained by direct fitting of the
experimental data using the favored HLS model
in Ref.~\cite{vmd}{} \cite{krmnew}.  
This indicates that, at least in the isospin
conserving case, matching of an ansatz of the type employed in
Ref.~\cite{ijl} to the OPE representation of the vector current
correlator provides a good fit to the actual vector current
spectral function.  Note that the matching between the hadronic
and OPE sides of the conventional QCD sum rule obtained in
Ref.~\cite{ijl} is also considerably better than that obtained
in earlier analyses using the narrow width approximation for the
$\rho$, further indicating the necessity of employing an ansatz
of the IJL form.  The only improvement in the IJL ansatz one
might have hoped for was the use of a variable $\rho$ width,
to be fixed by the sum rule analysis.  Since, however, in the
isospin conserving case, the result of such an exercise is
to essentially reproduce the width employed by IJL, it seems
unlikely that the analysis would have been significantly altered
by such an extension.

In light of the discussion above, we should thus be able
to extract the desired isospin breaking vector meson decay constants
from the results of Ref.~\cite{ijl}.  The results we use below,
however, differ somewhat from those quoted in that reference, and
the reasons for this difference must first be explained.
The first point in need of clarification is related to the
fact that there are two sets of 
results associated with the analysis using the physical
$\rho$ width in Ref.~\cite{ijl}.  Of these two sets, only the one contained in
the column labelled ``physical widths'' and ``no constraint'' in
Table 1 of that reference should be employed.  The reason for rejecting
the other set (labelled ``constrained'') is that these results were
obtained assuming $\Pi^{38}(m_\omega^2)$ could be extracted
from the $\omega\rightarrow\pi\pi$ contribution to the physical
cross section.  This assumption, however, as we have seen above,
neglects the presence of additional $\Pi^{38}$ 
contributions in the $\rho$ exchange
contribution to the cross-section, and hence is incorrect.
The second modification we make in employing the results of
Ref.~\cite{ijl} concerns the magnitude of the errors quoted
on the extracted values of the parameters $f_V$.  It turns
out that, in Ref.~\cite{ijl} an overly conservative error was
assumed on the crucial input isospin breaking light quark
mass ratio, $r_m=m_u/m_d$.  The authors of Ref.~\cite{ijl}
employed $r_m=0.50\pm 0.25$.  The ratio, $r_m$, however,
is actually much better constrained 
than this from ChPT analyses\cite{leutwylermasses}.  As a
consequence, the quoted errors in Ref.~\cite{ijl} are unnecessarily
inflated.  The authors of Ref.~\cite{ijl} have kindly provided unpublished
results corresponding to the more realistic input $r_m=0.54\pm 0.04$
employed in earlier sum rule analyses of the correlator at hand.
We will, therefore, employ, for determining the errors on the
extraction of the
desired isospin breaking vector meson decay constants, the results
of the ``unconstrained'' fit obtained using the modified input above,
$r_m=0.54\pm 0.04$.  The results, which correspond to a stable
Borel regime $1.15\ GeV<M<2.45\ GeV$, are\cite{derek97}
\begin{eqnarray}
f_\rho&=&4\sqrt{3}
F^{(3)}_\rho F^{(8)}_\rho =0.0030\pm 0.0012\ GeV^2\nonumber \\
f_\omega&=&-4\sqrt{3}
F^{(3)}_\omega F^{(8)}_\omega =0.0025\pm 0.0009\ GeV^2\nonumber \\
f_\phi&=&4\sqrt{3}F^{(3)}_\phi F^{(8)}_\phi =  -0.0002\pm 0.0002\ GeV^2.
\label{ijlresults}
\end{eqnarray}
The contributions from the $\rho^\prime$ and $\omega^\prime$ are
found to be very small, and cannot be reliably extracted from
the sum rule in its current form.  Note that the reliability
of the results is supported by that fact that they display
two features which correspond
to our general expectations:  first,
that $f_\rho\simeq f_\omega$ and, second, that $f_\phi$ is of order
$6$--$7\%$ of $f_\rho$ and $f_\omega$.

We are now in a position to evaluate the isospin breaking decay
constants $\ferho$, $\ftomega$ and $\ftphi$.  To do so we employ
the results of Eq.~(\ref{ijlresults}), together with the relations
\begin{equation}
F^{EM}_V=F^3_V+{\frac{1}{\sqrt{3}}}F^8_V\ ,
\label{femphysical}
\end{equation}
where the physical EM vector meson decay constants, $F^{EM}_V$,
determined experimentally from the partial widths for the
decays $V\rightarrow e^+e^-$, are
\begin{eqnarray}
F^{EM}_\rho &=&154\pm 3.6\ MeV\nonumber \\
F^{EM}_\omega &=&45.9\pm 0.8 \ MeV\nonumber \\
F^{EM}_\phi &=& -79.1\pm 2.3\ MeV .
\label{expfv}
\end{eqnarray}
The sign of the $\phi$ decay constant in Eq.~(\ref{expfv}) has
been chosen to be consistent with expectations from $SU(3)_F$
symmetry and ideal mixing.  Note that the relative magnitudes in
Eqs.~(\ref{expfv}) 
are roughly in line with the expectations of that limit (in which
one would expect the $\rho$, $\omega$ and $\phi$ decay constants
to be in the ratios $3:1:-\sqrt{2}$).  It is then straightforward
to solve for $F^{(3)}_V$ and $F^{(8)}_V$ separately.  
One finds, for the isospin breaking decay constants,
\begin{eqnarray}
F^{(8)}_\rho&=& 2.8\pm 1.1\ MeV \nonumber \\
F^{(3)}_\omega&=&-4.2\pm 1.5\ MeV \nonumber \\
F^{(3)}_\phi&=&0.21\pm 0.21\ MeV,
\label{isobreakingf}
\end{eqnarray}
where the quoted errors are totally dominated by the errors
of the sum rule fit values of the products of decay constants.
The values of the isospin conserving decay constants then
follow immediately from Eqs.~(\ref{femphysical}), (\ref{expfv}) and 
(\ref{isobreakingf}).  We will quote them in the form 
of ratios to the relevant experimental values, which form allows one to
directly compute the additional corrections required to obtain the pure
isovector and isoscalar spectral functions from those obtained
conventionally, i.e., via the standard analysis described above.
We find
\begin{eqnarray}
\left[ {\frac{F^{(3)}_\rho}{F^{EM}_\rho}}\right] -1 &=& 
-0.011\pm 0.0043 \nonumber \\
\left[ {\frac{F^{(8)}_\omega}{\sqrt{3}F^{EM}_\omega}}\right] -1 &=&
0.091\pm 0.029 \nonumber \\
\left[ {\frac{F^{(8)}_\phi}{\sqrt{3}F^{EM}_\phi}}\right] -1 &=&
0.0027\pm 0.0027\ .
\label{isoconservingf}
\end{eqnarray}

\section{Consequences for the Isovector and Isoscalar Spectral Functions}

If we drop terms second order in isospin breaking, then the $\rho$,
$\omega$ and $\phi$
resonance contributions to the isovector and isoscalar vector current
spectral functions are easily seen to be
\begin{eqnarray}
\left[ \rho^{33}(q^2)\right]_{V} &=& \left[ F^{(3)}_\rho\right]^2
\delta_\rho (s) \nonumber \\
\left[ \rho^{88}(q^2)\right]_{V} &=& \left[ F^{(8)}_\omega\right]^2
\delta_\omega (s)\, +\,  \left[ F^{(8)}_\phi\right]^2
\delta_\phi (s)\ .
\label{spectralreal}
\end{eqnarray}
The results of the standard extractions, in contrast, are obtained
by replacing $F^{(3)}_\rho$ with $F^{EM}_\rho$, $ F^{(8)}_\omega$
with $\sqrt{3}F^{EM}_\omega$ and $ F^{(8)}_\phi$ with
$\sqrt{3}F^{EM}_\phi$.  The corrections to be applied to the
standard contributions in order produce the true resonance contributions
to the isovector and isoscalar spectral functions are then given
by the ratios
\begin{eqnarray}
\left[ \frac{F^{(3)}_\rho}{F^{EM}_\rho}\right]^2&=& 0.979\pm 0.0086
\nonumber \\
\left[ \frac{F^{(8)}_\omega}{\sqrt{3}\, F^{EM}_\omega}\right]^2&=& 
1.189\pm 0.065 \nonumber \\
\left[ \frac{F^{(8)}_\phi}{\sqrt{3}\, F^{EM}_\phi}\right]^2&=& 
1.0054\pm 0.0054\ .
\label{corrections}
\end{eqnarray}

From the above results we see that the standard procedure leads
to an overestimate of the vector spectral function by $2.1\pm 0.9\, \%$.
This is still noticeably smaller than the $\sim 5\%$ errors on
the {\rxn} cross-sections in the resonance region.  As a result,
it is not yet possible, when comparing to
$\tau$ data, to see the effect of the $\rho^{38}$
contributions to the {\rxn} spectral functions extracted using
the conventional analysis.  (See Fig. 10c of Ref.~\cite{aleph97}
for a comparison of the spectral functions as extracted from
$\tau$ and $e^+e^-$ data.  The above correction would lower
the $e^+e^-$ points by $\sim 25$ nanobarns in the region of
the $\rho$ peak.)  A correction of this size, however, would 
certainly become important if one wished to make tests of CVC
at the $1\%$ level.  

A much greater surprise is the size of the correction required
in the case of the $\omega$ contribution to the isoscalar
spectral function.  While a $19\%$ isospin breaking correction
might sound unnaturally large, the size of the correction is, in
fact, completely natural, and easily understood.  The main 
features of the result follow from considering only the
leading order contributions as discussed in section II above.
Let us, therefore, consider the approximation
in which one considers only the leading (${\cal O}(m_q,q^0)$) 
contributions to $\rho$--$\omega$
mixing, neglects the ``direct'' isospin violating contributions
to the vector meson decay constants (which are also higher
order), and works in the 
ideal mixing/$SU(3)_F$ approximation in which the EM decay constant
of the pure isospin $1$ component of the $\rho$ is $3$
times that of the pure isospin $0$ component of the $\omega$.
Writing
\begin{equation}
\rho = \rho_I +\epsilon\, \omega_I \, ,\qquad
\omega =\omega_I -\epsilon\, \rho_I\ ,
\end{equation}
where the subscript $I$ denotes the isospin pure states, and
$\epsilon$ is ${\cal O}(\delta m)$, the physical EM decay constants
become
\begin{eqnarray}
F^{EM}_\rho&=&F^I_\rho +\epsilon F^I_\omega
\simeq F^I_\rho\left( 1+\frac{\epsilon}{3}\right) \nonumber \\
F^{EM}_\omega&=&F^I_\omega -\epsilon F^I_\rho
\simeq F^I_\omega\left( 1-3\epsilon \right) \, .
\end{eqnarray}
The fractional correction in the $\omega$ case is thus expected
to be $\sim 9$ times as big as that for the $\rho$ case.  That
the actual corrections turn out to be exactly a factor of
$9$ different is a numerical accident, but the large relative
size of the corrections is completely natural, and associated
with the smallness of the EM $\omega$ coupling and the
pattern of mixing in the vector meson sector.  Note also
the fact that the corrections are of opposite signs is 
exactly what one expects based on the general arguments above.

It is not just the isovector spectral function, with its relation
to CVC, for which the corrections obtained above are of 
phenomenological interest.
The difference of the isovector and isoscalar spectral functions also
enters a number of interesting sum rules, and these sum rules must,
therefore, also be corrected for the effects just discussed.
As an example we will consider the extraction of the
sixth (chiral) order low energy constant (LEC), $Q(\mu )$, (in the
notation of Refs.~\cite{gk96,km96}) from the inverse moment
chiral sum rule\cite{gk96}
\begin{eqnarray}
&&\int_{4m_\pi^2}^\infty  ds \ \frac{(\rho^{33}-\rho^{88})(s)}{s}
=
{\frac{16 (m_K^2-m_\pi^2)}{3F^2}} Q (\mu^2)
+ {\frac{1}{48 \pi^2}} \log \left( {\frac{m_K^2}{m_\pi^2}}\right) \nonumber \\
&&\qquad\qquad 
+\left(\frac{L_9^r(\mu^2)+L_{10}^r(\mu^2)}
{2 \pi^2 F^2}\right) \left[ m_\pi^2 \log \left({\frac{m_\pi^2}{\mu^2}}\right) 
- m_K^2\log\left({\frac{m_K^2}{\mu^2}}\right)\right]\ .
\label{invchmom}
\end{eqnarray}
In Eq.~(\ref{invchmom}), $L_k^r$ are the scale-dependent renormalized
fourth order LEC's of Gasser and Letuwyler\cite{gl85}, and $\mu$ is
the ChPT renormalization scale.  The form of this equation relies on
the two-loop expressions for the $33$ and $88$ correlators obtained
in Ref.~\cite{gk95}.
The difference of the $33$ and $88$ spectral functions also
enters a method of determining the strange current quark mass originally
suggested by Narison\cite{narisonqm}.  In this application,
a weighted integral over $\rho^{33}(q^2)-\rho^{88}(q^2)$ is performed, 
the weight function being that which enters inclusive $\tau$ decays.
The corrections in this case, and the resulting values of the
strange quark mass, will be treated in a separate paper\cite{km97new}.

In the analysis of the sum rule Eq.~(\ref{invchmom}) performed in
Ref.~\cite{gk96}, the $\rho$ contribution to the spectral integral
was obtained from $\tau$ decay data, and hence does not require
the correction discussed above.  The $\omega$ and $\phi$ contributions,
however, are determined from the experimental 
$V\rightarrow e^+e^-$ partial widths, and
hence contain contributions from $\rho^{38}$ which must be removed.
The uncorrected $\rho$, $\omega$ and $\phi$ contributions to the
spectral integral are\cite{gk96} $0.0374$, $-0.0103$ and $-0.0204$,
respectively.  Implementing the corrections above, one finds that
the sum of these three contributions is reduced from $0.0067$ to
$0.0046\pm 0.0007$, a downward shift of $31\%$.  Including the
estimates for the $4\pi$ and $\bar{K}K\pi\pi$ contributions
as evaluated in Ref.~\cite{gk96}, we find that the central value
for $Q(m_\rho^2)$ is shifted from $3.7\times 10^{-5}$ to
$2.2\times 10^{-5}$, a change of $41\%$.  As noted above, because
of the cancellations inherent in forming $\rho^{33}(q^2)-\rho^{88}(q^2)$ 
(the combination vanishes in the $SU(3)_F$ limit), the effect of
the isospin breaking corrections is large.  A similar effect is
found in the case of the strange quark mass analysis.

\section{Summary}
We have shown that contributions to {\rxn} involving an intermediate
state $\rho$ or intermediate state $\omega$ or $\phi$ contain
contributions from the isospin violating $38$ vector spectral function
which are not negligible, and must be removed if one wishes
to extract the isovector $33$ and isoscalar $88$ spectral functions
from {\rxn} data.  Using the results of a QCD sum rule analysis of
the $38$ correlator, we have been able to estimate the isospin
violating vector meson decay constants required to make these
subtractions.  We find that the isovector spectral function is
$\sim 2\%$ smaller than what one would obtain by assuming it
was identical to the full experimental $\rho$ contribution, and that the
$\omega$ contribution to the isoscalar spectral function is $\sim 19\%$
larger than what one would obtain from experiment without making
this correction.  We have also explained why it is unavoidable that
(1) the isoscalar correction will be much larger 
than the isovector correction (by roughly an order of magnitude), 
and (2) the sign of the
$\rho$ and $\omega$ corrections in the isovector and isoscalar cases,
respectively, will be opposite.  A consequence of the second point
is that all observables related to weighted integrals over the
difference of the $33$ and $88$ spectral functions will receive
large isospin breaking corrections, dominated by those which need
to be made to correctly obtain the $\omega$ contribution to
the $88$ isoscalar term.

Finally, we note that it might be possible to reduce the errors on
the extractions of the isospin breaking decay constants by updating
the sum rule analysis of the $38$ correlator using recent
improved values
for the input parameters, and evaluating higher order $\alpha_s$ corrections
to the Wilson coefficients appearing in the D=2,4 terms
of the OPE.  This will be the
subject of future investigations.
\acknowledgements
The author would like to thank Andreas H\"ocher of the ALEPH
collaboration for elaborating on the procedure used in making
isospin breaking corrections to {\rxn} data in the CVC test of
Ref.~\cite{aleph97}, and Derek Leinweber for making available the
unpublished results employed in Section III above.
The ongoing support of the Natural Sciences and
Engineering Research Council of Canada, and the hospitality of the
Special Research Centre for the Subatomic Structure of Matter at the
University of Adelaide, where this work was performed, are also
gratefully acknowledged.

\end{document}